# Integrated Fabry-Perot cavities as a mechanism for enhancing micro-ring resonator performance

Jiayang Wu,[1] Tania Moein,[1] Xingyuan Xu,[1] Guanghui Ren,[2] Arnan Mitchell,[2] and David J. Moss[1,a]

[1]*Centre for Micro-Photonics, Swinburne University of Technology, Hawthorn, Victoria 3122, Australia*

[2]*Electronics and Telecommunications Engineering, School of Engineering, RMIT University, Melbourne, Victoria 3001, Australia*

We propose and experimentally demonstrate the enhancement in the filtering quality (Q) factor of an integrated micro-ring resonator (MRR) by embedding it in an integrated Fabry-Perot (FP) cavity formed by cascaded Sagnac loop reflectors (SLRs). By utilizing coherent interference within the FP cavity to reshape the transmission spectrum of the MRR, both the Q factor and the extinction ratio (ER) can be significantly improved. The device is theoretically analyzed, and practically fabricated on a silicon-on-insulator (SOI) wafer. Experimental results show that up to 11-times improvement in Q factor, together with an 8-dB increase in ER, can be achieved via our proposed method. The impact of varying structural parameters on the device performance is also investigated and verified by the measured spectra of the fabricated devices with different structural parameters.

## I. INTRODUCTION

Along with the development of micro/nano fabrication technologies, integrated resonators with compact footprints, mass-producibility, high scalability, and versatile applications have come of age and become key building blocks for photonic integrated circuits.[1,2] The quality (Q) factor, defined as the ratio of resonance wavelength to full width at half maximum (FWHM),[1] is one of the fundamental parameters for integrated resonators. Integrated resonators with high Q factors are highly desirable and used for a wide range of applications such as narrow-bandwidth filters,[3,4] high-performance lasers,[5,6] high-efficiency nonlinear-optic devices,[7–13] and high-sensitivity sensors.[14,15]

To implement integrated resonators with high Q factors, a number of approaches have been proposed and demonstrated.[16–24] Since low internal cavity loss is a requirement to achieve a high Q factor, many attempts have been made to reduce the cavity loss, including modifying the fabrication process to reduce scattering loss induced by sidewall roughness,[16,17] employing high-confinement waveguides made from low-index materials,[7,18] designing resonant cavities with ultrasmall mode volumes,[4,15] and so forth.[3,6] Other methods based on mode interactions within coupled resonant cavities have also been investigated, including Fano[19–21] resonance and electromagnetically-induced-transparency (EIT) analogue[22–24] based methods. The Fano resonance based methods come at the expense of yielding sharp asymmetric filtering spectra for high Q factors, which have undesired distortions on the filtered signal and hence limit the applications of these filters. The EIT analogue based methods can achieve high-Q transmission peaks within the resonance notches, but the stopbands are usually limited by the linewidths of the resonance notches.

In this work, we propose and experimentally demonstrate a novel scheme to improve the Q factor of integrated resonators based on a novel device configuration employing an integrated Fabry-Perot (FP) cavity. By introducing a FP cavity, formed by cascaded Sagnac loop reflectors (SLRs), to reshape the transmission spectrum of an add-drop micro-ring resonator (MRR), the Q factor can be significantly increased, together with an increase in extinction ratio (ER). Unlike the approaches used to improve the Q factors based on asymmetric Fano resonances,[19–21] the filtering shape resulting from our approach can remain symmetric. Moreover, the bandwidths of the stopbands are no longer limited by the resonance linewidths as those in EIT analogue based approaches.[22–24] Our operation principle is universal, which can also apply to enhancing Q factors of other types of integrated resonators. We present a theoretical analysis for the operation principle, and fabricate the designed devices on a silicon-on-insulator (SOI) wafer using silicon fabrication technologies. A Q factor enhancement of 11 times and an ER improvement of 8 dB are obtained in the measured spectrum, which show good agreement with theory and confirm the effectiveness of our proposed method. The comparison between the measured spectra of different fabricated devices also verifies the impact of varying structural parameters on the device performance.

---

[a]Author to whom correspondence should be addressed. Electronic mail: dmoss@swin.edu.au

## II. DEVICE CONFIGURATION AND OPERATION PRINCIPLE

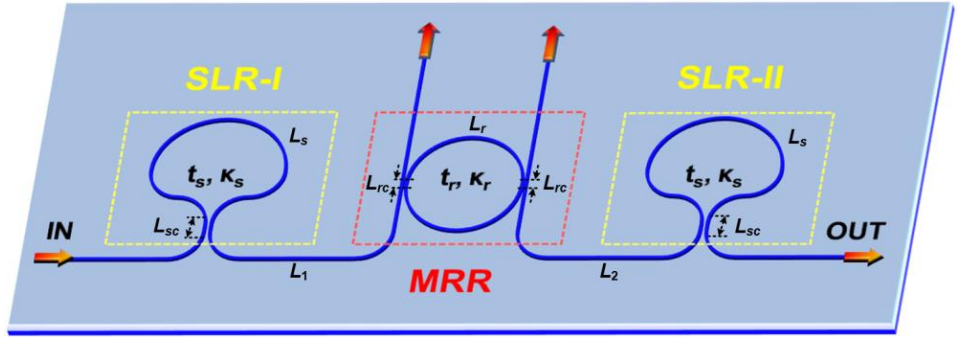

FIG. 1. Schematic configuration of the FP-cavity-assisted MRR (FP-MRR) filter with an add-drop MRR sandwiched between two Sagnac loop reflectors (SLR-I and SLR-II).

Figure 1 illustrates the schematic configuration of the FP-cavity-assisted MRR (FP-MRR) filter, which consists of an add-drop MRR sandwiched between a pair of SLRs (SLR-I and SLR-II). The cascaded SLR-I and SLR-II form a FP-cavity, which we term as SLR-FP cavity. The SLR-FP cavity reshapes the filtering spectrum of the MRR. Due to the sharpening of the filtering shape by the coherent interference within the SLR-FP cavity, the Q of the FP-MRR filter can be effectively increased. The definitions of the waveguide and coupler parameters of the FP-MRR filter are listed in Table I. Based on the scattering matrix method,[25] the field transmission function from port IN to port OUT is:

$$T_{\text{FP-MRR}} = \frac{T_{\text{MRR}} T_{\text{SLR}}^2 a_1 a_2 e^{i(\varphi_1+\varphi_2)}}{1 - T_{\text{MRR}}^2 R_{\text{SLR}}^2 a_1^2 a_2^2 e^{2i(\varphi_1+\varphi_2)}} \quad . \quad (1)$$

Note that $a_1\exp(i\varphi_1)$ and $a_2\exp(i\varphi_2)$ contribute equally to $T_{\text{FP-MRR}}$, so we use $L_1 = L_2$ in the design. $T_{\text{MRR}}$ in Eq. (1) denotes the field transmission function of the drop output of the MRR given by:

$$T_{\text{MRR}} = \frac{-\kappa_r^2 \sqrt{a_r}\ e^{i(\varphi_r/2)}}{1 - t_r^2 a_r e^{i\varphi_r}} \quad . \quad (2)$$

For identical SLR-I and SLR-II, $T_{\text{SLR}}$ and $R_{\text{SLR}}$ in Eq. (1) are the field transmission and reflection functions, respectively, which can be written as:

$$T_{\text{SLR}} = (t_s^2 - \kappa_s^2) a_s e^{i\varphi_s}, \quad (3)$$
$$R_{\text{SLR}} = -2it_s \kappa_s a_s\ e^{i\varphi_s}. \quad (4)$$

**TABLE I. Definitions of waveguide and coupler parameters in the FP-MRR filter**

| Waveguide | Length | Transmission factor[a] | Phase shift[b] |
|---|---|---|---|
| waveguide connecting SLR-I to MRR | $L_1$ | $a_1$ | $\varphi_1$ |
| waveguide connecting MRR to SLR-II | $L_2$ | $a_2$ | $\varphi_2$ |
| Sagnac loops in SLR-I and SLR-II | $L_s$ | $a_s$ | $\varphi_s$ |
| micro-ring in MRR | $L_r$ | $a_r$ | $\varphi_r$ |
| Coupler | Coupling length[c] | Field transmission coefficient | Field coupling coefficient |
| couplers in SLR-I and SLR-II | $L_{sc}$ | $t_s$ | $\kappa_s$ |
| couplers in MRR | $L_{rc}$ | $t_r$ | $\kappa_r$ |

[a] $a_i = exp(-\alpha L_i/2)$ ($i = 1, 2, s, r$), $\alpha$ is the power propagation loss factor.

[b] $\varphi_i = 2\pi n_g L_i/\lambda$ ($i = 1, 2, s, r$), $n_g$ is the group index, $\lambda$ is the wavelength.

[c] $L_{sc}$ and $L_{rc}$ are the straight coupling lengths shown in Fig.1. They are included in $L_s$ and $L_r$, respectively

The field transmission function for the SLR-FP cavity formed by cascaded SLR-I and SLR-II can be expressed as:

$$T_{\text{SLR-FP}} = \frac{T_{\text{SLR}}^2 a_1 a_2 e^{i(\varphi_1 + \varphi_2)}}{1 - R_{\text{SLR}}^2 a_1^2 a_2^2 e^{2i(\varphi_1 + \varphi_2)}} \quad . \tag{5}$$

Based on Eqs. (1) – (4), the calculated power transmission spectrum of the FP-MRR filter ($|T_{\text{FP-MRR}}|^2$) is depicted in Fig. 2(a). The structural parameters are chosen as follows: the radius of the microring in the MRR is $R = 20$ μm, $L_s = 129.66$ μm, and $L_1 = L_2 = 77.67$ μm. The gap size of all the directional couplers is 120 nm, and the straight coupling lengths are $L_{sc} = 2$ μm and $L_{rc} = 0.5$ μm. For single-mode silicon photonic nanowire waveguides with a cross-section of 500 nm × 220 nm, the calculated field transmission coefficients using Lumerical MODE Solutions are $t_s = \sim0.92$ and $t_r = \sim0.95$ accordingly. We also assume that the waveguide group index of the transverse electric (TE) mode is $n_g = 4.3350$ and the power propagation loss factor is $\alpha = 101$ m$^{-1}$ (4.3 dB/cm), which are based on our previously fabricated devices with the same structural parameters.[26] The calculated power transmission spectra of the MRR ($|T_{\text{MRR}}|^2$), the SLR-FP cavity ($|T_{\text{SLR-FP}}|^2$), and the cascaded MRR and SLR-FP cavity ($|T_{\text{Cascaded}}|^2 = |T_{\text{MRR}} \times T_{\text{SLR-FP}}|^2$) are also presented in Fig. 2(a). It is clear that both Q factor and ER of the proposed FP-MRR filter are significantly increased as compared with those of the MRR and the SLR-FP cavity. The comparison between $|T_{\text{FP-MRR}}|^2$ and $|T_{\text{Cascaded}}|^2$ also indicates that the increase in Q factor is due to more than just cascaded filtering. Although there is a slight increase in the insertion loss (IL), it is within a reasonable range. The Q factor and IL of $|T_{\text{FP-MRR}}|^2$ in Fig. 2(a) are $3.3742 \times 10^4$ and 1.91 dB, respectively. For a single add-drop MRR that has the same Q-factor, the corresponding $t_r$ is 0.9869 and the IL is ~1.86 dB. The calculated $|T_{\text{FP-MRR}}|^2$ for various values of $t_s$ are shown in Fig. 2(b). There is Q factor enhancement in the FP-MRR filter as compared with the MRR for each value of $t_s$. To quantify the enhancement of Q factor, we further define Q enhancement factors (QEFs) of the FP-

MRR filter and the cascaded MRR and SLR-FP cavity as $QEF_{FP-MRR} = Q_{FP-MRR} / Q_{MRR}$ and $QEF_{Cascaded} = Q_{Cascaded} / Q_{MRR}$, respectively, with $Q_{FP-MRR}$, $Q_{MRR}$, and $Q_{Cascaded}$ denoting the Q factors of $|T_{FP-MRR}|^2$, $|T_{MRR}|^2$, and $|T_{Cascaded}|^2$ accordingly. The calculated $QEF_{FP-MRR}$, $QEF_{Cascaded}$, and IL of the FP-MRR filter for various values of $t_s$ are plotted in Fig. 2(c). One can see that $QEF_{FP-MRR}$ increases when $t_s$ approaches $\sqrt{1/2}$, together with an increase in IL. This is because the SLRs work as total reflectors with $T_{SLR} = 0$ when $t_s = \sqrt{1/2}$. The comparison between $QEF_{FP-MRR}$ and $QEF_{Cascaded}$ reveals that the FP-MRR filter exhibits better performance in Q factor enhancement. The calculated $|T_{FP-MRR}|^2$ for various values of $t_r$ are shown in Fig. 2(d). One can see that increasing $t_r$ enhances Q-factor and ER, although at the expense of an increase in IL.

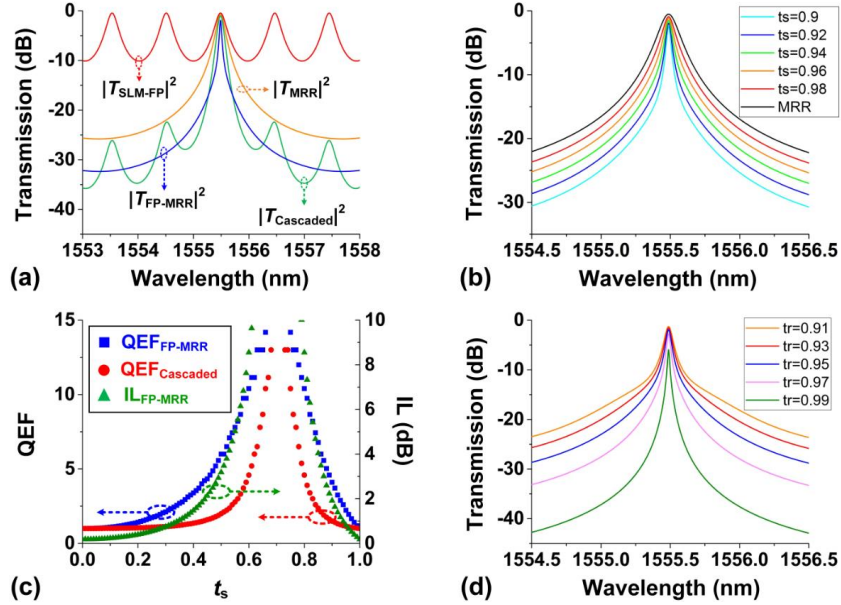

FIG. 2. (a) Calculated power transmission spectra of the FP-MRR filter ($|T_{FP-MRR}|^2$), the MRR ($|T_{MRR}|^2$), the SLR-FP cavity ($|T_{SLR-FP}|^2$), and the cascaded MRR and SLR-FP cavity ($|T_{Cascaded}|^2$). (b) Calculated $|T_{FP-MRR}|^2$ for various $t_s$ as compared with $|T_{MRR}|^2$. (c) Calculated $QEF_{FP-MRR}$ (blue squares), $QEF_{Cascaded}$ (red circles), and IL of the FP-MRR filter (olive triangles) for various $t_s$. (d) Calculated $|T_{FP-MRR}|^2$ for various $t_r$.

It should be noted that the lengths $L_{1,2}$ in Figs. 2(a) – (d) are designed to satisfy:

$$L_{SLR-FP} = (N + 0.5) L_r, (N = 0, 1, 2, \ldots) \tag{6}$$

where $L_{SLR-FP} = 2(L_1 + L_2 + L_s)$ is the round-trip length of the SLR-FP cavity. If Eq. (6) is not satisfied, it may lead to an asymmetry in the filtering shape, as shown in Fig. 3(a). $\Delta\varphi$ represents the phase detuning along $L_1$, and $\Delta\varphi = 0$ corresponds to the condition that Eq. (6) is satisfied. There is also an obvious improvement in Q factor and ER for the asymmetric filtering shape. When $\Delta\varphi = \pi/2$, i.e., $L_{SLR-FP} = NL_r$ ($N = 0, 1, 2, \ldots$), the power transmission spectrum of the FP-MRR filter exhibits high-order symmetric filtering shape.[27] When Eq. (6) is satisfied, the increase of $N$ in Eq. (6) leads to an enhanced Q factor of the FP-MRR, as shown in Fig. 3(b). Such enhancement is not obvious unless the change of $N$ is large enough. When $N$ continues to increase until the free spectral range (FSR) of the SLR-FP cavity approaches the linewidth of the MRR, filtering shapes with multiple transmission peaks begin to appear. It should also be pointed out that the FP-MRR is a universal approach to improving filter performance that works not only for MRR but, by changing the configuration of the FP or replacing the MRR, it can also apply to other integrated resonators such as microdisks and photonic crystal cavities.

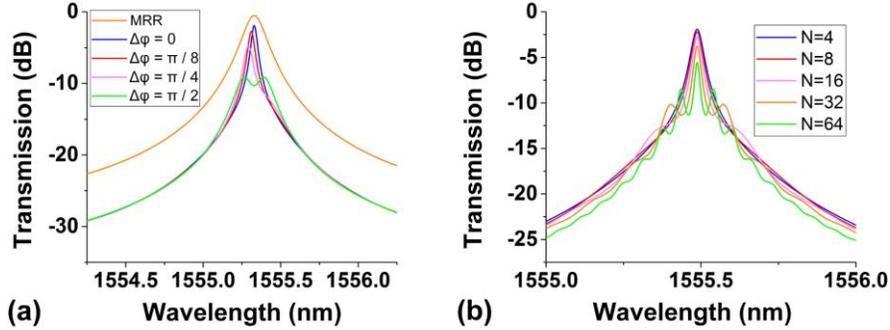

FIG. 3. (a) Calculated $|T_{FP\text{-}MRR}|^2$ for various $\Delta\varphi$ along $L_1$ as compared with $|T_{MRR}|^2$. (b) Calculated $|T_{FP\text{-}MRR}|^2$ for various $N$ in Eq. (6).

## III. DEVICE FABRICATION

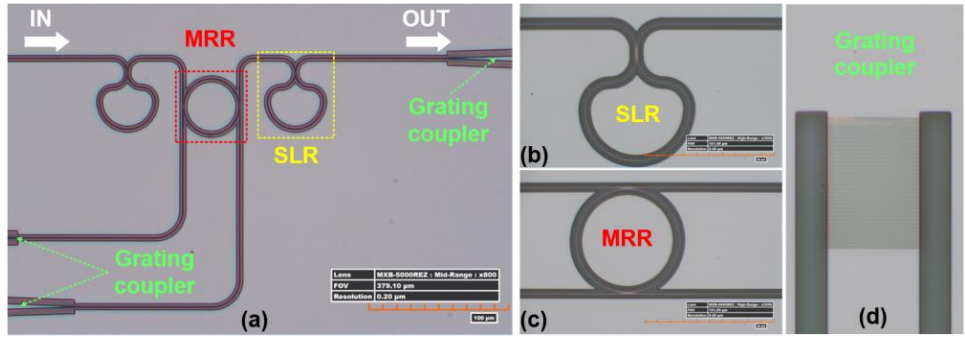

FIG. 4. (a) Micrograph for one of the fabricated FP-MRR filters on a SOI wafer. (b) – (d) Zoom-in micrographs of the SLR, the MRR, and the grating coupler, respectively.

The FP-MRR filters based on the above principle were fabricated on a SOI wafer with a 220-nm-thick top silicon layer and a 3-µm-thick buried oxide (BOX) layer. The device fabrication involves standard complementary metal-oxide-semiconductor (CMOS) processes only, with the exception that the device pattern is defined using electron-beam lithography. The micrographs for one of the fabricated devices are shown in Figs. 4(a)–(d). In our fabrication, electron beam lithography (Vistec EBPG 5200) was utilized to define the device layout on positive photoresist (ZEP520A), followed by a reactive ion etching (RIE) process to transfer the device pattern to the top silicon layer. During the RIE process, $SF_6$ and $CHF_3$ were used as the etching gases. Grating couplers for TE polarization that were 70-nm shallow-etched were employed at the ends of port IN and port OUT to couple light into and out of the chip with single mode fibres, respectively. The grating couplers were fabricated by a second electron beam lithography step, together with another RIE process. Gold markers, prepared by metal lift-off after photolithography and electron beam evaporation, were used for alignment. The unused ports of the MRR were also connected to grating couplers to dissipate undesired signal with negligible reflection. For each fabricated FP-MRR filter, we also fabricated a MRR and a SLR-FP cavity, which had the same size as those of the add-drop MRR and the SLR-FP cavity in the FP-MRR filter, respectively. This was done to facilitate a comparison of Q factor and ER in the measured transmission spectra.

## IV. DEVICE CHARACTERIZATION AND ANALYSIS

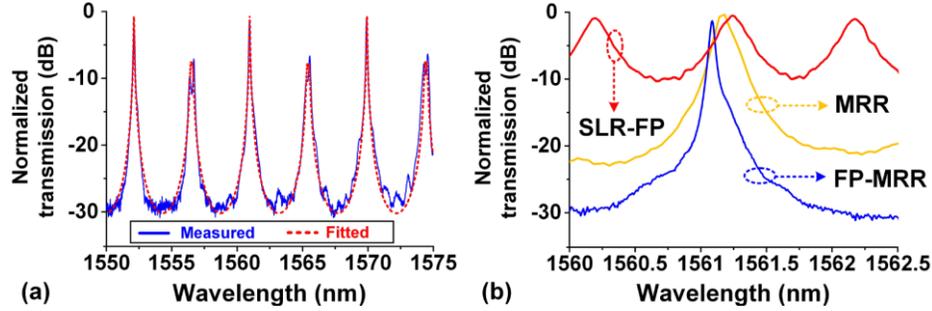

FIG. 5. (a) Measured (solid curve) and fitted (dashed curve) transmission spectra of the fabricated FP-MRR filter. (b) Zoom-in spectrum of (a) around one resonance at ~1561 nm, as compared with measured transmission spectra of the fabricated MRR and SLR-FP cavity with the same size as those of the FP-MRR filter in (a).

The normalized transmission spectrum for one of the fabricated devices is shown in Fig. 5(a) by the blue solid curve. The spectrum is measured by a fast-scanned continuous-wave (CW) laser (Keysight 81608A) and recorded by a high-sensitivity optical power meter (Keysight N7744A) at an input optical power of ~0 dBm. The grating coupler loss is ~4.6 dB each, or ~9.2 dB for both, which is subtracted from the measured spectrum. The normalized spectrum is then fit by the red dashed curve calculated from Eqs. (1) – (4). The fitting parameters are $t_s$ = ~0.9224, $t_r$ = ~0.9413, $n_g$ = ~4.3262, and $\alpha$ = ~121 m$^{-1}$ (~5.3 dB/cm), which are close to our expectations before fabrication. It can be seen that the experimentally measured curve agrees well with the theory. The resonances with single peaks and the resonances with split peaks appear alternately. The generation of the resonances with split peaks is due to the coincidence of the transmission peak of the MRR and the transmission valley of the SLR-FP, which is determined by the condition in Eq. (6). Due to the Vernier-like effect[28] between the SLR-FP and MRR, the FSR of the FP-MRR is determined by the MRR. Therefore, the Q factor can be enhanced without sacrificing the FSR. Since the finesse of a resonator is defined as the ratio of the FSR to the resonance FWHM[1], the FP-MRR is superior to the method of improving Q by simply increasing the resonant cavity length in terms of finesse enhancement. Given that the nonlinear phase within a resonator is proportional to the square of the finesse[29], enhanced nonlinear effect in the FP-MRR can also be anticipated. Figure 5(b) illustrates a zoom-in spectrum of the fabricated FP-MRR filter around one resonance at ~1561 nm, together with the measured transmission spectra of the fabricated MRR and SLR-FP cavity with the same size as those of the FP-MRR filter. It is clear that the linewidth of the fabricated FP-MRR filter is much lower than those of the fabricated MRR and the SLR-FP cavity. The Q factor of the fabricated FP-MRR filter is ~28,000, which is ~3.4 and ~4.3 times as high as that of the fabricated MRR and the SLR-FP cavity, respectively. There is also a 7-dB improvement in the ER of the fabricated FP-MRR filter as compared with that of the fabricated MRR. The IL of the FP-MRR filter is ~1.8 dB, which is 0.7 dB higher than that of the MRR. A comparison of the three measured transmission spectra in Fig. 5(b) confirms Q factor and ER improvements in the fabricated FP-MRR filter. Note that there are slight shifts in resonance wavelengths of the three curves in Fig. 5(b), which is mainly attributed to fabrication error between different devices. The filtering shape of the fabricated FP-MRR filter shows a slight asymmetry, and this is because the desired condition in Eq. (6) is not perfectly satisfied. By introducing thermo-optic micro-heaters[30, 31] or carrier-injection electrodes[32] along $L_{1,2}$ to tune the phase shift, the symmetry of the filtering shape can be further improved.

To demonstrate the impact of $t_s$, $t_r$, and $\Delta\varphi$ on the filtering shape of the proposed FP-MRR filter, we also fabricated a number of devices with different structural parameters. The measured transmission spectra of the fabricated FP-MRR filters with $L_{sc}$ = 1, 2, 3 µm are presented in Fig. 6(a). The resonance wavelengths of different devices are normalized for comparison. The length of $L_{1,2}$ is changed accordingly to compensate the difference in $L_{sc}$ and keep the same $L_{SLR-FP}$, thereby the devices still operate around the desired condition in Eq. (6). In Fig. 6(a), increasing $L_{sc}$ leads to an enhanced coupling strength, which corresponds to a decreased $t_s$ in Fig. 2(b). For $L_{sc}$ = 3 µm, the Q factor of the fabricated FP-MRR filter is as large as ~90,000, which is ~3 times as high as that of the device with $L_{sc}$ = 2 µm and ~11 times as high as that of the fabricated MRR. The ER and IL for $L_{sc}$ = 3 µm is ~29.1 dB and ~7.2 dB, respectively, which are ~8.0 dB and 6.1 dB higher than those of the fabricated MRR. The measured Q factors and ILs of the fabricated FP-MRR filters with various $L_{sc}$ are shown in Fig. 6(b). For continuously increased $L_{sc}$, i.e., continuously decreased $t_s$, the Q factor and IL of the fabricated FP-MRR filter first increase and then decrease, which shows good agreement with the theory in Fig. 2(c). Since we have demonstrated in Ref. 33 that dynamic tuning of $t_s$ can be realized by using interferometric couplers to replace the directional couplers and tuning them in a differential mode, tuning the Q factor of the proposed FP-MRR filter can also be achieved in the same way. Figure 6(c) depicts the measured transmission spectra of the fabricated FP-MRR filters with various $L_{rc}$, which correspond to different values of $t_r$. The Q factor and ER of the fabricated FP-MRR filter increase with decreasing $L_{rc}$, i.e., increasing $t_r$, together with an increase in IL. This agrees well with the theory in Fig. 2(d). The measured transmission spectra of the fabricated FP-MRR filters with various $L_1$ are shown in Fig. 6(d), which correspond to different values of $\Delta\varphi$. It can be seen that asymmetric filtering shapes similar to those in Fig. 3(a) are obtained, which are also consistent with our theoretical prediction.

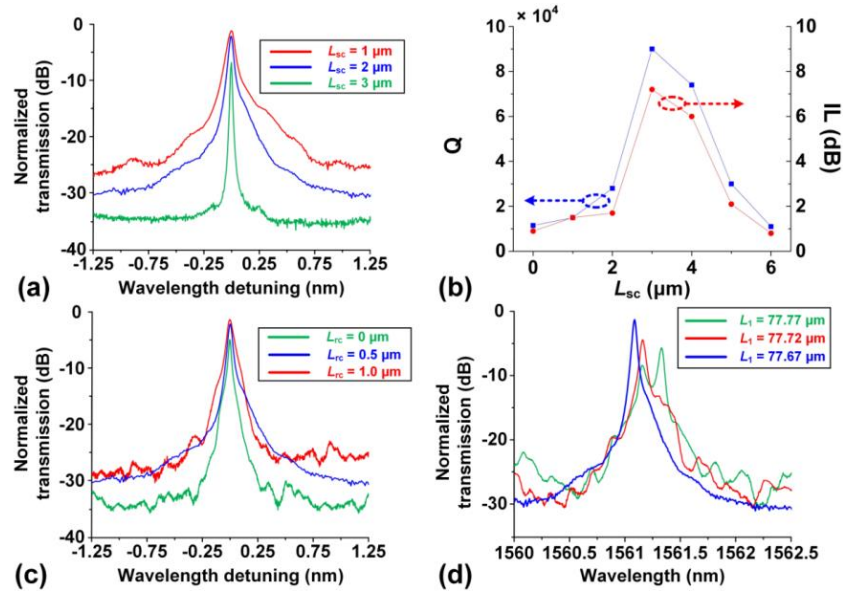

FIG. 6. (a) Measured transmission spectra of the fabricated FP-MRR filters with various $L_{sc}$. (b) Measured Q factors and ILs of the fabricated FP-MRR filters with various $L_{sc}$. (c) – (d) Measured transmission spectra of the fabricated FP-MRR filters with various $L_{rc}$ and $L_1$, respectively.

## V. CONCLUSION

In summary, we propose and demonstrate the Q factor enhancement of MRR filters through the use of an integrated FP cavity formed by cascaded SLRs. We achieve up to 11-times enhancement in Q factor together with an 8-dB increase in ER for a device fabricated on a SOI wafer. Our approach relies on coherent interference within the FP cavity which sharpens the filtering shape of the MRR. We investigate the impact of varying different parameters on device performance, and our experimental results agree with theory. These results validate our approach as an effective way to enhance the Q factor of integrated MRRs.

## ACKNOWLEDGMENTS

This work was supported by the Australian Research Council Discovery Projects Program (DP150104327). This work was performed in part at the Melbourne Centre for Nanofabrication (MCN) in the Victorian Node of the Australian National Fabrication Facility (ANFF).